\documentclass[12pt]{article}
\usepackage{epsfig}
\begin{document}

\centerline{\bf\Large{The flatness of the Universe is robust}}

\vspace*{1.00 cm}

\centerline{Matts Roos \& S. M. Harun-or-Rashid} 
\centerline{Department of Physics, Division of High Energy Physics} 
\centerline{University of Helsinki, Finland.} 

\vspace*{1.00 cm}


The flatness of the Universe is measured by the combination of 
parameters $\Omega_0 =\Omega_m + \Omega_{\Lambda}$. Here $\Omega_m$ is  
the dimensionless density parameter of gravitating matter comprising 
baryons, neutrinos and some yet unknown kinds of dark matter, 
$\Omega_{\Lambda}$ is the density parameter of vacuum energy, related 
to the cosmological constant $\Lambda$ by $\Omega_{\Lambda} = 
\Lambda/3H^2_0$, and $H_0$ is the Hubble constant, usually given in the 
form $H_0 = 100h$ km s$^{-1}$ Mpc$^{-1}$. A flat universe is defined 
by the condition $\Omega_0 =1$.

There are many observations contributing information on $\Omega_0$. 
Although some results may be affected by unknown systematic errors 
rendering published errors of specified significance dubious, it is 
still of interest to know what combined value they determine for 
$\Omega_0$ at present. In a sequence of reports 
(Roos \& Harun-or-Rashid 1998, 1999, 2000) we have combined all 
published observations quoting a value and an error in a least-squares 
fit assuming all the errors to be Gaussian. This method excludes the use 
of many interesting analyses quoting only limits. 
 
Before the advent of the two very recent balloon experiments 
BOOMERANG (de Bernardis \& al. 2000, Lange \& al. 2000) and MAXIMA-1 
(Balbi \& al. 2000) our most recent fit 
(for details see Roos \& Harun-or-Rashid 2000) used 9 constraints: 
parametrizations of earlier CMBR data (Lineweaver 1998, Tegmark 1999) 
marginalized to the 
$(\Omega_m ,\Omega_{\Lambda})$-plane, the cluster mass function 
combined with the linear mass power spectrum in Ly$\alpha$ data 
(Weinberg \& al. 1999), X-ray cluster evolution data (Bahcall, Fan 
and Cen 1997, Eke \& al. 1998), constraints from 14 classical double 
radio galaxies (Daly, Guerra \& Wan Lin 1998), the SN Ia data 
(Riess et al. 1998, Perlmutter et al. 1998, 1999), and analyses of the 
power-spectrum of extragalactic objects (Broadhurst \& Jaffe 1999, 
Roukema \& Mamon 1999). All these constraints represent different 
functions of $\Omega_m$ and $\Omega_{\Lambda}$, not functions simply 
of $\Omega_0$. 

The result was then 
\begin{eqnarray}
\Omega_0 = \Omega_m + \Omega_{\Lambda}=0.94\pm 0.14\ 
\end{eqnarray}
(or $\pm 0.21$ for the two-dimensional fit).

The total $\chi^2$ was 4.1 for 7 degrees of freedom, much too low for 
statistically distributed data. From this we can conclude that the 
various errors quoted for the nine constraints are not statistical: 
they have been blown up unreasonably by added systematic errors, and 
there is no motivation for blowing them up further by adding more 
arbitrary systematic errors (as is often advocated by the referees of 
our papers). The true statistical error should be of the order of 
0.15 to 0.18.
 
Let us now turn to the new balloon data. BOOMERANG observes 
(de Bernardis \& al. 2000) that the position of the first multipole 
peak occurs at $\ell = 197\pm 6$ which corresponds to 
\begin{eqnarray}
\Omega_0 = (200/\ell )^2 = 1.03\pm 0.06\ .
\end{eqnarray}

This value is very weakly dependent on a larg number 
of parameters which mostly get determined by the shape of the multipole 
spectrum above the region of the first peak (Lange \& al. 2000).

Let us make a statistical comment to the above BOOMERANG value. It has 
already been hailed in the literature as a proof that the Universe is 
closed. But nobody can escape statistical fluctuations (which do not 
reflect the quality of an experiment), therefore it is prudent to say 
that the true value of the peak is unknown, but with 68\% probability 
it lies in the range between 0.97 and 1.09.

It should therefore not have come as a surprise that the MAXIMA-1 
experiment (Balbi \& al. 2000) reported $\Omega_0 = 0.90\pm 0.15$,
where the error corresponds to a 95\% confidence range. We convert 
this to a 68\% error, thus
\begin{eqnarray}
\Omega_0 = 0.90\pm 0.08\ .
\end{eqnarray}

Averaging the values (1), (2) and (3) we find  
\begin{eqnarray}
\Omega_0 = 0.97\pm 0.05\ .
\end{eqnarray}

The two balloon experiments have clearly reduced the combined error 
dramatically, but the central value is robust, almost unchanged. One 
concludes that the Universe is flat to a very good precision, and one 
can take the robustness as a good indication of the trustability of 
the simple $\chi^2$ method.


\section*{References}
Bahcall N.A., Fan X., Cen R., 1997, ApJ 485, L53 \\
Balbi A., et al., 2000, astro-ph/0005124 \\
Broadhurst T., Jaffe A.H., astro-ph/9904348 \\
Daly R.A., Guerra E.J., Wan Lin, astro-ph/9803265  \\
de Bernardis P., et al., 2000, Nature 404, 955 \\
Eke V.R. et al., 1998, MNRAS 298, 1145  \\
Lange A.E., et al., 2000, astro-ph/0005004 \\
Lineweaver Ch.H., 1998, ApJ 505, L69  \\
Perlmutter S et al., 1998, Nature 391, 51; 1999, ApJ 517, 565 \\
Riess A.G. et al., 1998, AJ 116, 1009  \\
Roos M., Harun-or-Rashid S.M., 1998, A\&A 329, L17  \\
Roos M., Harun-or-Rashid S.M., 1999, Proc. HEP'99, IOP Publishing  \\
Roos M., Harun-or-Rashid S.M., 2000, astro-ph/0003040  \\
Roukema B.F., Mamon G.A., 1999, astro-ph/9911413  \\
Tegmark M., 1999, ApJ 514, L69  \\
Weinberg D.H. et al., 1999 ApJ 522, 563 \\

\label{lastpage}

\end{document}